\documentclass[3p, twocolumn]{elsarticle}
\usepackage[utf8]{inputenc}
\usepackage{amssymb}
\usepackage{amsmath}
\usepackage{amsthm}
\usepackage{cleveref}
\usepackage{mhchem}
\usepackage{lineno}
\usepackage{subcaption}
\usepackage{appendix}
\usepackage[colorinlistoftodos]{todonotes}
\usepackage{url}

\usepackage[T1]{fontenc}
\usepackage{lmodern}

\theoremstyle{definition}

\newcommand{\pety}{\ensuremath{P_{\left(\left[E, t\right],Y\right)}\left(\left[\varepsilon, \tau\right],y\right)}}
\newcommand{\pey}{\ensuremath{P_{\left(E,Y\right)}\left(\varepsilon,y\right)}}
\newcommand{\pet}{\ensuremath{P_{\left[E, t\right]}\left(\left[\varepsilon, \tau\right]\right)}}

\newcommand{\tildepety}{\ensuremath{\tilde{P}_{\left(\left[E, t\right],Y\right)}\left(\left[\varepsilon, \tau\right],y\right)}}
\newcommand{\tildepet}{\ensuremath{\tilde{P}_{\left[E, t\right]}\left(\left[\varepsilon, \tau\right]\right)}}
\newcommand{\Iota}{\ensuremath{I}}
\newcommand{\unit}[1]{\ensuremath{\;\mathrm{#1}}}
\newcommand{\csfour}{\ensuremath{\ce{^{134}Cs}}}
\newcommand{\csseven}{\ensuremath{\ce{^{137}Cs}}}

\newcommand{\kforty}{\ensuremath{\ce{^{40}K}}}
\newcommand{\ththirtytwo}{\ensuremath{\ce{^{232}Th}}}
\newcommand{\ueight}{\ensuremath{\ce{^{238}U}}}
\newcommand{\ufive}{\ensuremath{\ce{^{235}U}}}
\newcommand{\gfour}{\texttt{geant4}}

\title{Improvement of Nuclide Detection through Graph Spectroscopic Analysis Framework and its Application to Nuclear Facility Upset Detection}
\author{Pedro Rodríguez Fernández, Christian Svinth, Alex Hagen}
\date{June 2025}

\begin{document}
\begin{abstract}
    We present a method to improve the detection limit for radionuclides using spectroscopic radiation detectors and the arrival time of each detected radiation quantum.  We enable this method using a neural network with an attention mechanism. We illustrate the method on the detection of Cesium release from a nuclear facility during an upset, and our method shows $2\times$ improvement over the traditional spectroscopic method. We hypothesize that our method achieves this performance increase by modulating its detection probability by the overall rate of probable detections, specifically by adapting detection thresholds based on temporal event distributions and local spectral features, and show evidence to this effect. We believe this method is applicable broadly and may be more successful for radionuclides with more complicated decay chains than Cesium; we also note that our method can generalize beyond the addition of arrival time and could integrate other data about each detection event, such as pulse quality, location in detector, or even combining the energy and time from detections in different detectors.
\end{abstract}
\maketitle

\section{Motivation}

Radiation spectroscopy is a key pillar in nuclear material detection and analysis; its applications range from nuclear safety and public health to national security. Its key tenet is that the energy of emitted radiation from a material is unique to that material. Unfortunately, nuclear, detector, and absorber effects convolve the energy of emitted radiation with nuisance variables. This tends to obscure the discriminative signatures from each material.  Increasingly advanced detection equipment and data analytic methods have been employed to attempt to detect discriminative signatures at lower quantities or to measure them to lower quantification uncertainty with the same counting time.

This work's novel contribution is to extend the definition of a ``discriminative signature'' in radiation spectroscopy beyond that of energy - to include time of detection, species of detected particle, and conceivably other information.  This extension continues an existing line of research in the radiation spectroscopy community which has already made strides in fields such as coincidence spectroscopy and track reconstruction.  We generalize the definition sufficiently to allow for flexible applications, and provide a framework with which to apply our definition in realistic situations.  We profile our framework's performance on an important task in the field: the detection of cesium release from a nuclear facility, and show improvements in detection performance stemming from our novel framework.

\section{Background}\label{sec:background}

In this work, we consider the detection and spectroscopy of radiation quanta such as the electron (beta), helium nucleus (alpha), photon (gamma), or neutron. In a generalized radiation detector model such as that presented by Knoll \cite{Knoll2003}, the detector generates electrical current after irradiation by said quanta.  One can attempt to measure the current generated by a single quantum with a suitably responsive electrical circuit; under this model the effects generated by the quanta will appear as electrical pulses which can be measured, digitized, and analyzed by suitable electronics.  The measurement and analysis of such pulses is collectively known as pulse height spectroscopy \cite{Knoll2003}, hereafter simply spectroscopy.

The magnitude of the pulse in the detection electronics depends on many factors, including the energy of the quanta when emitted, the geometry and composition of any intervening material between the point of emission and the detector, the detector material, voltage, temperature, and condition, and settings within the measurement electronics. Hypothetically, when all conditions above are held constant, a difference in emitted energy of a quantum can be detected. This hypothesis, coupled with a hypothesis that unique materials emit quanta of unique energies, laid the foundation for an extremely successful field of material detection and analysis.

The largest challenges of spectroscopy relate to the extent to which the above conditions can be held constant, and the extent to which the hypotheses are true.  It is remarkably difficult and restrictive to manufacture consistent detectors, to retain consistent electronics settings, and to ensure constant and consistent environmental settings and intervening material.  Beyond these realities, however, three unavoidable truths in spectroscopy conspire to convolve the detector response of different energies such that they may overlap in pulse height; those of detector resolution effects, those of the effects of the nuclear processes inherent in some quanta emission, and those of the effects of partial deposition of the quanta's energy in the detector.
\begin{description}
    \item[Resolution Effects] Radiation detectors have an inherent resolution, stemming from the movement of the detectors' constituent atoms. This manifests itself in the ``broadening'' of the distribution of measured pulse height when subjected to repeated quanta of the same energy. A simplistic model of this effect is to draw random numbers from a normal distribution centered at the quanta's true energy, with standard deviation some polynomial (usually order two) of that energy.
    \item[Nuclear Effects] Some radioactive emission processes do not have a single characteristic energy. For example, in beta emission, an antineutrino is also emitted. Therefore, the beta emission process into a three body problem: the beta, the antineutrino, and the recoiling nucleus which decayed.  The three constituents can then take different amounts of energy, each. In consequence, beta decaying isotopes are identified by a ``beta spectrum'' defined by an endpoint energy - below which the emission of a beta bearing any energy is possible to different extents.
    \item[Partial Deposition Effects] The quanta may not fully deposit all its energy in the radiation detector. If a quanta only partially interacts within the detector, the pulse measured will have commensurately less pulse height, and therefore seem indicative of a lower energy quanta.
\end{description}

Therefore, the field of radiation spectroscopy has been primarily concerned with the ability to infer the emitted energy of detected quanta, and further to infer the identity of the source of that quanta.  Advanced analytics have been developed and have proven remarkably successful. Radiation spectroscopists can identify constituents of materials in the parts per million (for an example see \cite{ppm_1}) and in more advanced cases in the parts per billion (for an example see \cite{ppb_1}). 
The goal of this work is to extend and improve these these advanced analytics by developing a framework which focuses not only on the detected energy of the quanta, but other available information as well.

\subsection{Counting Experiment Analysis}

The framework developed in this work is inspired by modern information theory, and represents an attempt to unify the fields of information theory and radiation spectroscopy. Common analysis of spectroscopic data can be classified as a \emph{counting experiment}, where multiple discrete events are counted and their number and properties are used collectively to infer the presence or quantity of materials.  In the most basic counting experiment, two measurements are taken for equal time. One of these measurements includes the presence of an item of interest, the other does not.  Pulses are enumerated for each measurement, and if these are statistically significantly different, that material emits radiation.

Spectroscopy extends the basic concept of this counting experiment. Two measurements of equal time are taken, and pulses are again enumerated, but this time in a \emph{conditional} sense: i.e. only pulses are enumerated in a given energy window. In this way, one can identify a material that exhibits emission of a certain energy, even if it is of low activity compared to the background but has a unique energy of emission.

Quantification in spectroscopy can also be done in this conditional way.  When pulses are enumerated in a given energy window, the (radio-)activity of the material can be computed by utilizing the branching ratio of the material for that specific decay mode and the efficiency of the detector at that pulse height. When appropriate error propagation is performed, this can provide incredibly high precision results.

It is also possible to condition the counting experiment on additional information, as the additional information may qualify or disqualify pulses.  Many types of additional information are possible for this conditioning. In fact, traditional spectroscopy has been extended to use the rate of background emissions as additional information by counting only those pulses which fall in a certain energy window \emph{and are in excess of the baseline number of detections at that energy window}. Less common information such as pulse quality \cite{Hagen2022, Mace&Ward} has also been used. Most notable for this work is the conditioning of counting experiments based on the time of emission.

Coincidence and Anti-Coincidence spectroscopy are the best current examples of time conditioning on spectroscopy. Certain isotopes exhibit very short-lived states, in which they emit two radiation quanta within quick succession.  For a detector and electronics capable of measuring and recording the time and pulse height of each pulse (so-called ``list-mode'' data), these events can be reliably measured simply by disqualifying any events further than a small time window apart. Because of the relative rarity of detecting two random events during the same time window, the background rejection of this conditional counting experiment is superior to traditional spectroscopy, even in high activity samples or convolution regions of the spectrum.  Anti-coincidence spectroscopy is the converse, where pulses are disqualified because of their coincidence in time with a detection event \cite{Pierson2017, Pierson2022}. Coincidence spectroscopy is not limited to a single species of radiation quanta; for example beta- and gamma- coincidences are commonly used in radioxenon detection \cite{radio_coin1, radio_coin2, radio_coin3}. Recent extensions beyond two events \cite{triple_coincidence} have shown even more benefits.

The high energy physics community has taken the generalization of these discrete events even further.  In modern colliders, new physics are sought by using extremely high collision energies between subatomic particles.  Track reconstruction, the goal of analysis of these events, is to match Feynman diagrams of theoretically possible events to detection events in different detectors arrayed around the point of collision.  Track reconstruction can be viewed as a very generic version of the above conditional counting experiment: each count is qualified or disqualified by the detector in which it was detected (including the location of that detector), the time of the detection, the energy of the detection if applicable, and the structure between all detection events for a single collision.  Advanced classical and deep learning based machine learning techniques are applied, across many colliders \cite{Alonso-Monsalve2020, track1, track2, track3}.

\subsubsection{Classifier Based Counting Experiment}

The examples of advanced analysis of counting experiments in the previous section have been mainly developed through iterative development. A framework which can connect the physical processes and detection relevant figures of merit with the broader field of data analytics and science will benefit the overall development of new analytics. For this work, we use the Classifier Based Counting Experiment (CBCE) framework introduced by \cite{Hagen2022} to connect information theoretic tools with the detection relevant figures of merit.  In CBCE, a classifier - which is a foundational type of data analytic - scores each detection event on a scale of 0 to 1, indicating a score in how ``signal-like'' the event is. The classifier in a CBCE may range from the trivial (e.g. all events are considered signal), to the very complex (e.g. a trained neural network).  For different figures of merit and different background environments, different thresholds on that score can be chosen, above which an event is then classified as signal. Raw counting experiments, traditional spectroscopy, coincident spectroscopy, and even graph-based spectroscopy fit into this framework.

\subsubsection{Attention Mechanism}


The CBCE framework allows for the use of any classifier which accepts the relevant input data, and therefore is amenable to utilizing emerging deep learning networks for classification.
In this work we use a Transformer network \cite{Vaswani2017}, which utilizes the powerful \textit{Self-Attention} mechanism.
The self-attention mechanism allows the network to learn a relational matrix amongst all elements of the input sequence, identifying how different elements of the sequence interact with and influence each other, thereby producing an information-rich contextual embedding.
In the context of our work, this mechanism allows the network to perform comparisons and learn relationships amongst multiple detection events within the same detection window, thereby allowing us to modulate our predictions about a current detection event based on what events we have seen recently.

In this work, we specifically utilize \textit{Relative Global Attention} \cite{shaw2018self}.
Relative global attention differs from the standard attention mechanism in that it allows for weighting of sequence elements based on their relative positions to each other, rather than just their absolute position within the overall sequence.
In practice, we expect this relative positional information to be useful for identifying products which manifest as multiple detection events with a predictable time interval in between.

The mathematical formalization of our method for CBCE spectroscopy using list-mode data and advanced classifiers follows.

\section{Mathematical Formalism}

We consider a counting experiment where one or more detectors are used to detect discrete radiation quanta interactions within their bulk.  We detect $N$ discrete pulses; each with ``metadata'', including at least the instantaneous time of the detection $t$ and the detector in which it was detected $d$. The metadata also likely includes the pulse height of the event $h$, but could include additional event-wise information.  This constitutes a set of detections:
\begin{equation}
    V=\left\{v_{1}, v_{2},\;\dots, v_{N}\right\} 
\end{equation}
where $n$ represents the event along its metadata and $v_{n}$ is defined as
\begin{equation}
    v_{n} = \left[t_{n}, d_{n}, \dots\right]^{\top}\forall\;n\in\left[1, N\right].
\end{equation}

Some events in $V$ may be related to other events in $V$, such as the same nucleus decaying through different states. We can represent these relations as a set of tuples of datapoints of the form 
\begin{equation}
    E \equiv \{ \varepsilon_{m_{1}, m_{2}} = \left[v_{m_{1}}, v_{m_{2}}\right]^{\top} | \quad v_{m_{1}}, v_{m_{2}} \in V \}
\end{equation}
where $\varepsilon_{m_{1}, m_{2}}$ represents the connection between two events and, $m_{1}$ and $m_{2}$ represent the index of nodes in $V$ such that $v_{m_{1}}$ is a pulse related to $v_{m_{2}}$ and $t_{m_{1}} < t_{m_{2}}$. 

The above is a formulation of conditional counting experiments into a graph $G=(V, E)$ consisting of the nodes as pulses and their connections as edges. In radiation spectroscopy as described above, inference of the isotope of origin of each pulse in $V$ is desired.  That is, we would like to infer an additional property $\iota_{n}$ for each pulse in $V$. Hereafter, to align with traditional graph analytic nomenclature, we will refer to each pulse in $V$ as a vertex, and each connection in $E$ as an edge.

This framework can be applied retrospectively to all described applications of spectroscopy in \cref{sec:background}. For example, the original counting experiments can be described as statistical comparison of the cardinality of the graph generated in the blank and experiment; single detector spectroscopy is statistical comparison of the cardinality of subgraphs (conditional on the energy property of each vertex) of the graph generated by the blank and experiment; and coincident spectroscopy is the comparison of the cardinality of subgraphs of the blank and experiment, where the subgraphs are chosen by the presence of an edge between two vertices. Track reconstruction is the classification of the overall graph by various operations on $G$ (which may not require the inference of $\iota$ on the vertices).

It is clear from the above that radiation spectroscopy has predominantly used $V$ to infer $\iota$, with some limited uses of edges in $E$.  This is understandable: access to the ground truth of $E$ is currently impossible outside of simulation, and there are $\left|E\right| = \frac{\left|V\right|(\left|V\right|-1)}{2}$ possible edges in the graph which can be prohibitively large to infer. The first novel contribution of this work is the hypothesis that by inferring edges $E$, and using the entire graph for inference, better detection or quantification performance is possible.

Using information about each vertex, as well as information about the edges in the graph, we can create a classifier ($g$) which maps each vertex to a predicted isotope such that
\begin{equation}
    g : V \rightarrow \Iota
\end{equation}
where $\Iota$ is the set of all possible isotopes in the present situation. Again note that the classifier $g$ may be conditioned on the entire graph (to include predicted edges).

While we demonstrate an application as empirical evidence in support of our hypothesis above, we believe it to be an insight that is broadly applicable. In this section, we provide theoretical justification to motivate our framework's use in other detection problems.

\subsection{Justification}

We hypothesize that graph-based spectroscopy is better than traditional or even coincident spectroscopy in the general sense.  In order to prove that the new technique is ``better'', we must first define a Figure of Merit (FOM) under which a method of analysis can be measured to be better.  To define a figure of merit, we must first define the tasks of analysis of radiometric experiments.  We postulate that the most common and important tasks for which radiometric experiments are performed are:
\begin{description}
\item[Detection] the determination of whether a specific component (usually isotope) is present in the experiment.
\item[Quantification] the measurement of the (radio-)activity of a specific component in the experiment.
\item[Identification] the detection and quantification of those components which are present in the experiment. We note that this can be decomposed to be a combination of detection and quantification.
\end{description}

Many sources define two primary FOM for detection and quantification: the detection limit ($\ell_{d}$) and quantification uncertainty ($Q_{u}$). The detection limit is given by:

\begin{equation}\label{eq: ld}
    \ell_{d} = \frac{\ell_{c} + \frac{k^{2}}{2} + k\sqrt{\frac{k^{2}}{4} + \ell_{c}}}{\eta RT}
\end{equation}
where $\ell_{c}$, $\eta$ and $RT$ are defined as

\begin{equation}
    \begin{aligned}
        \ell_{c} &= 2.33\sqrt{ n_{fp}+0.4} + 1.35 \\
        \eta &= \frac{t_{p}}{t_{p} + 1 - t_{n}} \\
        RT &= n_{n}
    \end{aligned}
\end{equation}
and quantification uncertainty ($Q_{u}$)
\begin{equation}\label{eq: qu}
    Q_{u} = n_{p} + \frac{1}{A^{2}}\left[ n_{p}B + n_{n}C\right]
\end{equation}
where $A$, $B$ and $C$ are defined as 
\begin{equation}
    \begin{aligned}
        A &= t_{p}t_{n} - f_{p}f_{n} \\
        B &= t_{n}^{2}t_{p}(1-t_{p}) + f_{n}^{2}f_{p}(1-f_{p}) + 2t_{n}t_{p}f_{n}f_{p} \\
        C &=t_{n}^{2}f_{n}(1-f_{n}) + f_{n}^{2}t_{n}(1-t_{n}) + 2t_{n}^{2}f_{n}^{2}, \\
    \end{aligned}
\end{equation}
and $t_{p}$, $t_{n}$ are the normalized conditional accuracies where $f_{p} = 1- t_{n}$, $f_{n} = 1 - t_{p}$. We therefore adopt these two FOMs and we encourage to readers to refer to \cite{Hagen2021, Hagen2022, Knoll2003, MARLAP} for more information about them.

The FOM for identification is use-case dependent, but can be decomposed into its detection and quantification parts.

Next, we introduce the concept of a Classifier Based Counting Experiment (CBCE) from \cite{Hagen2022}.  In this concept, a classifier is used on each detection event to determine whether the event originated from the component of interest and therefore considered signal, or did not originate from the component of interest and therefore is considered background.  The results from \cite{Hagen2022} derived the detection limit and quantification uncertainty for any CBCE based on the conditional accuracies of the classifier, alongside physical considerations such as detector efficiency. The latter are invariant to the type of classifier used, and thus do not factor any further in comparisons between the methods.

The conditional accuracies then fully define the performance of CBCE methods for both detection and quantification.  Therefore, any classifier improvement which improves the conditional accuracies will necessarily improve the FOM. However, the improvement in conditional accuracies is classifier and experiment specific.  We seek not to prove that a specific classifier is better than another, but that a different set of input information (i.e. no time information in traditional spectroscopy, discrete time information in coincident spectroscopy, and continuous time information in graph-based spectroscopy) to the classifier results in a better CBCE and therefore detection or quantification method.

In general, the Mutual Information (MI) of the predictions from a classifier (hereafter "predictions") is a function of the conditional accuracies of the predictions \cite[p. 10]{Meyen2016} given by

\begin{equation}\label{eq: MI}
    \begin{aligned}
        MI= 1 - (w_{p}H_{2}(t_{p}) + w_{n}H_{2}(t_{n}))
    \end{aligned}
\end{equation}
where $H_{2}(p)$ ($p$ is the probability of occurrence) is the binary entropy given by 
\begin{equation}
    H_{2}(p) = p\log_{2}(\frac{1}{p}) + (1-p)\log_{2}(\frac{1}{1-p})
\end{equation}
and $w_{p}$ and $w_{n}$ are weights given by
\begin{equation}
    \begin{aligned}
        w_{p} &= \frac{n_{p}}{n_{p} + n_{n}} \\
        w_{n} &= \frac{n_{n}}{n_{p} + n_{n}}
    \end{aligned}
\end{equation}

Further, the mutual information between the predictions and the labels is maximally bounded by the mutual information between the observable features of the detection event (i.e. time, energy, and other measureables, hereafter "features") and the labels through the data processing inequality \cite{Beaudry2011}.

This result can be intuitively interpreted as providing two ways to improve the performance of a detection system through the CBCE framework.  The first is to improve the mutual information between the features and the labels, the second is to improve the mutual information between the predictions and the labels, given a fixed set of features. In Appendix \cref{sec:proof-feature-mi-increase}, we prove that adding information about detection events can never reduce the mutual information between the features and the labels. This strong result shows that, in the general case, adding information can give us a better chance at performing detection or quantification of isotopes.

\begin{figure}
    \centering
    \includegraphics{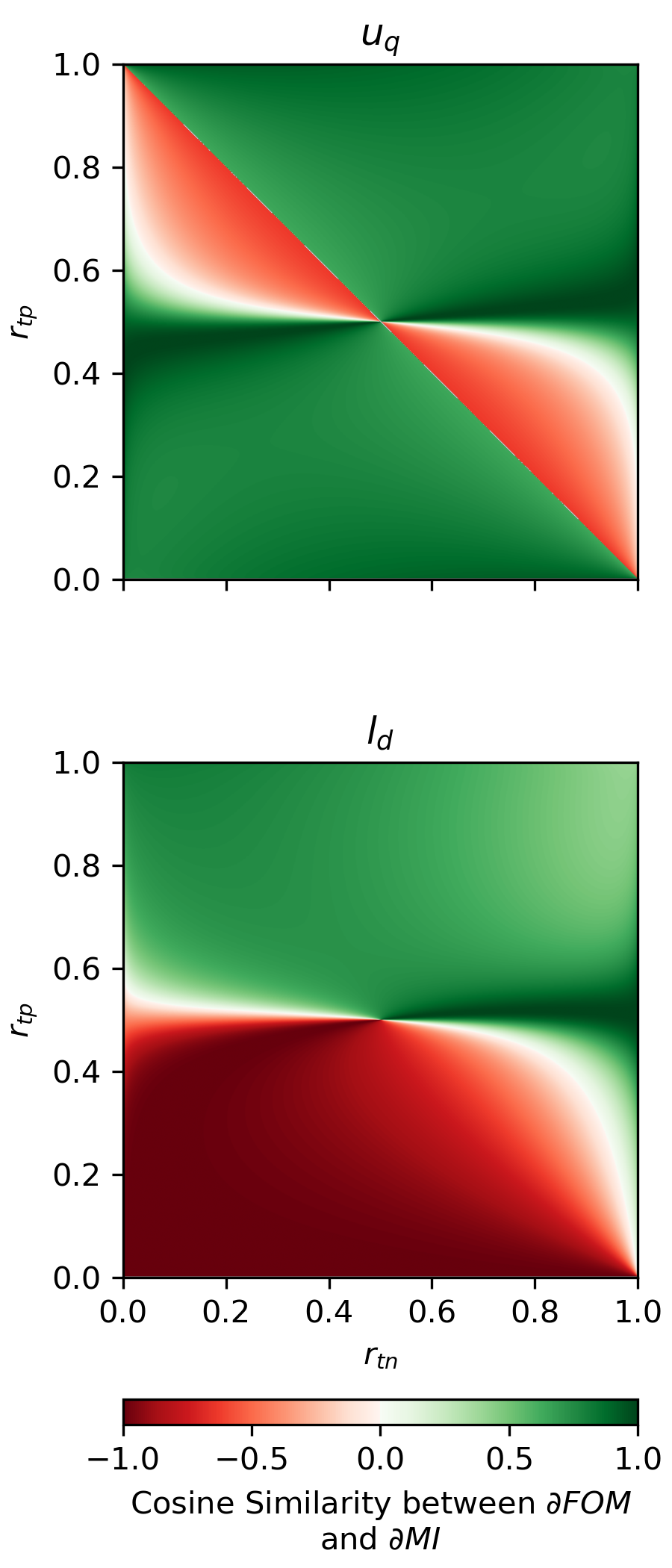}
    \caption{The cosine similarity between the direction of steepest ascent for MI and the direction of steepest descent for FOMs, assuming a number of observations of signal of $10,000$ and observations of background of $1,000$. Red areas indicate where the direction that would improve MI would actually reduce the performance against FOMs.}
    \label{fig:delta-fom}
\end{figure}

Interestingly, an improvement in MI does not always improve the detection limit or quantification uncertainty. This can be seen numerically in computing the derivatives of MI and the FOM with respect to the true positive and true negative rate (i.e. $\frac{\partial MI}{\partial \left[r_{tp},r_{tn}\right]^{T}}$ - hereafter abbreviated as $\partial MI$, $\frac{\partial l_{d}}{\partial \left[r_{tp},r_{tn}\right]^{T}}$ - hereafter abbreviated as $\partial l_{d}$, and $\frac{\partial u_{q}}{\partial \left[r_{tp},r_{tn}\right]^{T}}$ - hereafter abbreviated as $\partial u_{q}$). It can be seen in \cref{fig:delta-fom} that there are regions where the cosine similarity between these two vectors is negative, indicating that the direction of improvement for MI is opposite that of the FOM.  
While we have been unable to prove this analytically, we believe this is due to the ratio between the number of observed signal events and the number of observed background events. In calculating MI, this ratio is used to weight the entropy between the accuracy on each class, effectively removing the effect of any imbalance between the number of observations.  This is not the case for the figures of merit. We believe future work exploring the implications of these regions where MI improves and FOM degrades would be fruitful, as would the investigation of classification loss functions (which generally seek to improve MI) that improve FOM.

\subsection{Physical systems under which it is plausible}
In Appendix \cref{sec:tjp-must-vary}, we show that the relative likelihood of observing a detection from signal and background (the temporal joint probabilities (TPJs)) must \emph{vary} over time for an improvement to the detection figure of merit to be satisfied. Physical mechanisms under which this likelihood might vary would include:

\begin{description}
    \item[Radioactive Decay] Variation in the activity present due to radioactive decay.
    \item[Other Source Time Variation] Variation in source intensity due to other reasons, such as radiation generator settings, chopper movement or collimation movement.
    \item[Parent-Child Variation] Variation in the probability of observing a certain decay because of a prior decay. This is the generalization of the concept underlying coincidence spectroscopy: that given a previous observed decay, one can expect certain energies from that decay's daughters on a known time structure.
    \item[Position Variation] Variation in the spatial position of isotopes over time (e.g. migration, biological removal, atmospheric transport, or a combination of these mechanisms).
\end{description}

We must be able to \emph{observe} this variation, which serves to set some bounds on the above.  In general, the time process must change fast enough for its variation to be observed during the experiment, and slow enough to be resolved by the timing electronics of the system. Practical definitions of the limits of ``fast enough'' and ``slow enough'' is left for future work.

With the theoretical justification for using
\begin{itemize}
    \item As much information about each detection event as possible,
    \item Classifiers, including advanced classifiers such as graph-based neural networks, to classify each event as background or signal, and
    \item The CBCE framework to understand and optimize our overall detection analytic,
\end{itemize}
we now profile our techniques on the challenging problem of detection of Cesium release from a nuclear facility.

\section{Methods}

We envisioned a discrete air monitoring system to detect radioactive particulate effluents from a nuclear facility. As such, we assumed a small filter would have stack flow passed through it, and then removed from the stack and moved into a well detector within a count laboratory for monitoring. Particulates would be mainly similar to soil, thus our assumption of KUT background and the search for radioactive cesium particulates.

In order to generate a dataset to test the GSA framework, we created a simulation of Cesium release from a reactor and its putative detection at remote distances for a ground based high purity germanium detector.  We chose simulation to evaluate the performance of the framework, it was preferable to have ground-truth of the originating nucleus of each detection, which is most easily achieved by using simulation. Note that any ''list-mode`` dataset can be used with our framework; evaluation of the framework itself requires such ground-truth. This simulation provides time- and energy-resolved detection events for the inherent potassium, uranium, and thorium (KUT) commonly present in soil \cite{Keillor2018} and cesium release from a reactor.

The simulation of cesium release from a reactor upset is challenging and requires high fidelity modeling of isotopic ratios generated during the fission process, their migration within fuel, and physical details about the upset and release event. Activity ratios of $\ce{^{134}Cs}/\ce{^{137}Cs}$ measured from inside and within the nearby vicinity of the Fukushima Nuclear Power Plant after upset averaged around 1.0 \cite{Thakur2013}. While this number is indicative more of the specifics of the Fukushima upset than of any broad class of reactor upsets, we used this ratio in our simulations. A different ratio could be easily simulated, and our method is designed to be applicable to any distribution of signal nuclides, if they are present in the training data.

The simulation of KUT background from soil is also complicated to simulate.  Uranium and Thorium are primordial radionuclides \cite{primordial}, and therefore may have been born hundreds of millions of years ago. Monte Carlo simulation starting from birth of those isotopes would be computationally expensive, requiring many samples due to the low probability of an isotope born so long ago decaying during the relatively miniscule counting window.  The origin of potassium on earth is not as well understood, and is therefore even more of a challenge to simulate from first principles. To create a realistic simulation despite these uncertainties and challenges, we modify the approach in \cite{Keillor2018} because it is validated empirically against measurements.

In \cite{Keillor2018}, first the time to equilibrium for the decay chains from \kforty{}, \csseven{}, \ththirtytwo{}, \ueight{}, and \ufive{} are determined. We used the \texttt{radioactivedecay} \cite{radioactivedecay} library for its ease of integration with the rest of our toolset.  The time to equilibrium is $1\times10^{9}\unit{hr}$ for \kforty{}, $1.0\unit{hr}$ for \csseven{}, and $1\times10^{11}\unit{hr}$ for Th and U.
Then, the relative activity of isotopes starting at the beginning of the counting time is the product of the activity fraction at equilibrium and the proportion of each decay chain in soil. For the isotopic proportions, we used the values from DOE/RL-96-12 \cite{DOERL9612}, which are close to the values from \cite{Keillor2018}.

This process gives us the isotopic inventory of a cesium release from a reactor upset and the background inherent in soil; we still need to determine the response of a detector when those nuclides are present.  To do this, we developed a \gfour{} \cite{Agostinelli2003} simulation, based on the baseline \texttt{rdecay02} program provided with the framework.  In this simulation, a high purity germanium detector was simulated as a cylinder of germanium with radius $15\unit{cm}$ and length $7.5\unit{cm}$ with an internal cylindrical well of radius $0.5\unit{cm}$ and depth $4.1\unit{cm}$, inside a can of aluminum with thickness $1\unit{mm}$.  A cylinder of $50\unit{cm}$ radius and length $1.5\unit{m}$ of soil, comprised of the elements and proportions from \cite{Keillor2018} was modeled below the detector.  A source sample of radius $0.5\unit{cm}$ and thickness $0.1\unit{cm}$ was located in the bottom of the well. Air, using ratios of oxygen and nitrogen from the \texttt{rdecay02} baseline, was modeled as a cylinder of $50\unit{cm}$ and length $1.575\mathrm{m}$ above the soil, excluding the detector volume.  Events from the KUT nuclide inventory described above were generated uniformly throughout the top $25\unit{cm}$ of the soil\footnote{Uniformly distributing background decays throughout the entire volume of the soil would be more realistic, but more computationally intensive. We chose this variance reduction as this work focuses mainly on proving that a method works, not exact realism.} for background simulation.  Events from the Cs nuclide inventory and KUT nuclide inventory were generated uniformly throughout the source volume for sample simulation.
\begin{figure}
    \centering
    \includegraphics[width=\columnwidth]{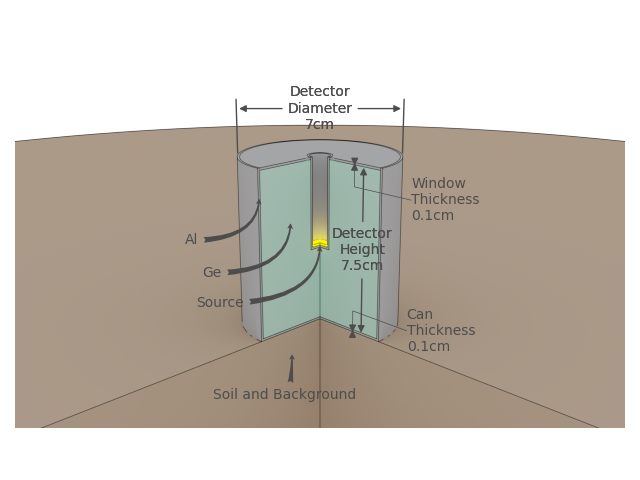}
    \caption{Three dimensional rendering of simulation geometry for detection of reactor upsets against background.}
    \label{fig:render}
\end{figure}

We utilized \gfour{}'s RadioactiveDecay module to determine the energy, species, and time of emissions from each isotope in the nuclide inventory. An event was started with zero energy of the specified radionuclide, and the time for each decay was artificially set to a time uniformly sampled within the count time window. Thereafter, \gfour{} simulates the further decay, transportation and interaction of the emitted particles, recoil nuclei and their daughters with the soil, ground, and detector. Each event is comprised of many tracks, which are comprised of all the transportation and interaction steps for every particle emitted during the decay and all its descendants. For those interactions and transportation steps happening after the end of the count time window, we stopped simulation of that track. For interactions in the detector, we capture any electron emission and record its full energy and time of interaction, stopping simulation of that track thereafter\footnote{Again, simulation of electron interactions within the detector after their emission would provide more realism, but we chose this variance reduction because the focus of this work is on analytic methodology}. The time, energy, and event unique identifier are saved for every energy deposition within the detector into a comma separated values file.  These energy deposition events are then collected into a binary object for easy loading into analysis programs thereafter.

\begin{figure*}
    \centering
    \includegraphics{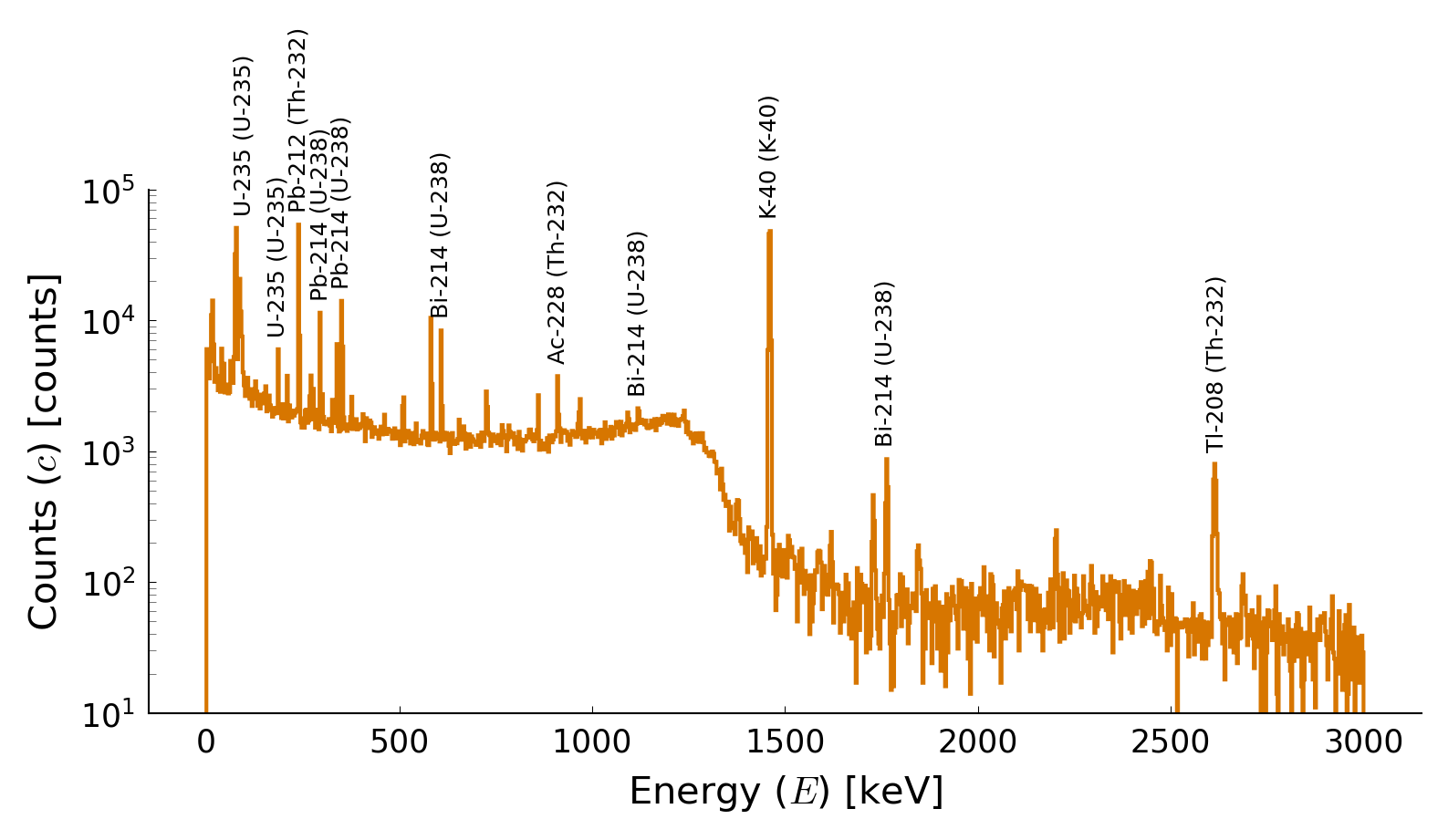}
    \caption{Spectra of detection events in an HPGe detector from KUT background using our simulation methodology modified from \cite{Keillor2018}, with selected peaks used for quantification of KUT from \cite{Heier1963, Abusini2008, Choi2018}.}
    \label{fig:spectra_vs_keillor}
\end{figure*}

\begin{figure}
    \centering
    \includegraphics[width=\columnwidth]{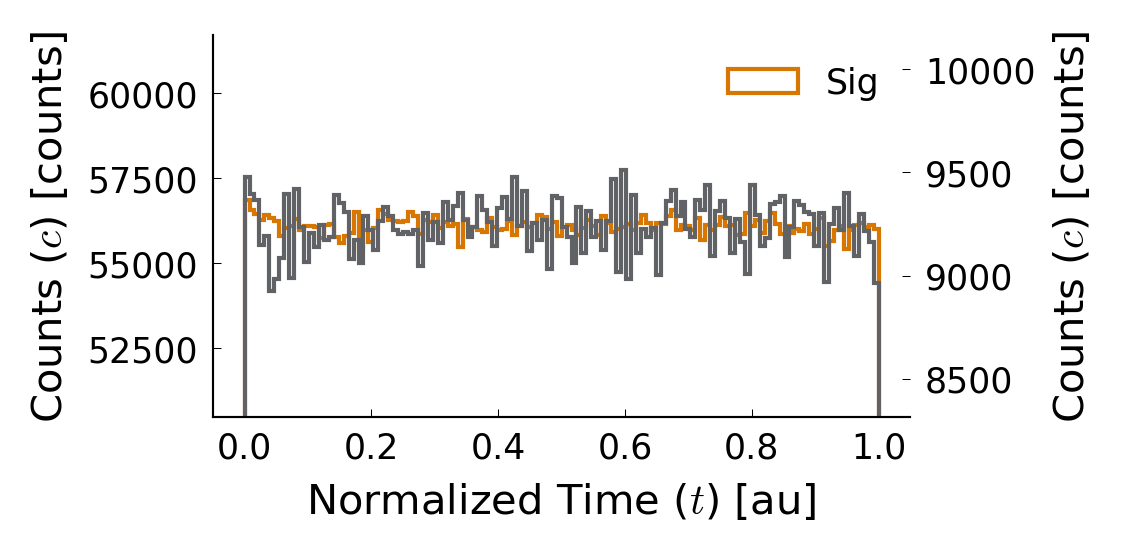}
    \caption{Timing characteristics of simulation of signal (Cs) and background (KUT) in the geometry from \cref{fig:render}. The emission time of all events in a simulation of X events from background and Y events from cesium release.}
    \label{fig:emission_time}
\end{figure}

This methodology provides an extension to \cite{Keillor2018} which provides realistic time resolution of each detection event.  Spectroscopically, the results of our simulation are very similar to \cite{Keillor2018}\footnote{The exact shape of our spectrum, especially the continuum and low energy features, are different than \cite{Keillor2018} because our detector geometry does not match their detector geometry.} and match peaks expected from literature on the quantification of KUT \cref{fig:spectra_vs_keillor}.  The emission time of all events is shown in \cref{fig:emission_time}, showing a uniform rate. This shows that some of the original isotopes' daughters also decay during the counting time, which is an example of extra information beyond energy that may make the proposed method useful.

Similar results were obtained for our signal simulation. Expected peaks which are used for the quantification of \csfour{} and \csseven{} \cite{Schoetzig1998} are present within the energy spectra, and the expected half lives are also present.

\begin{figure*}
    \centering
    \includegraphics{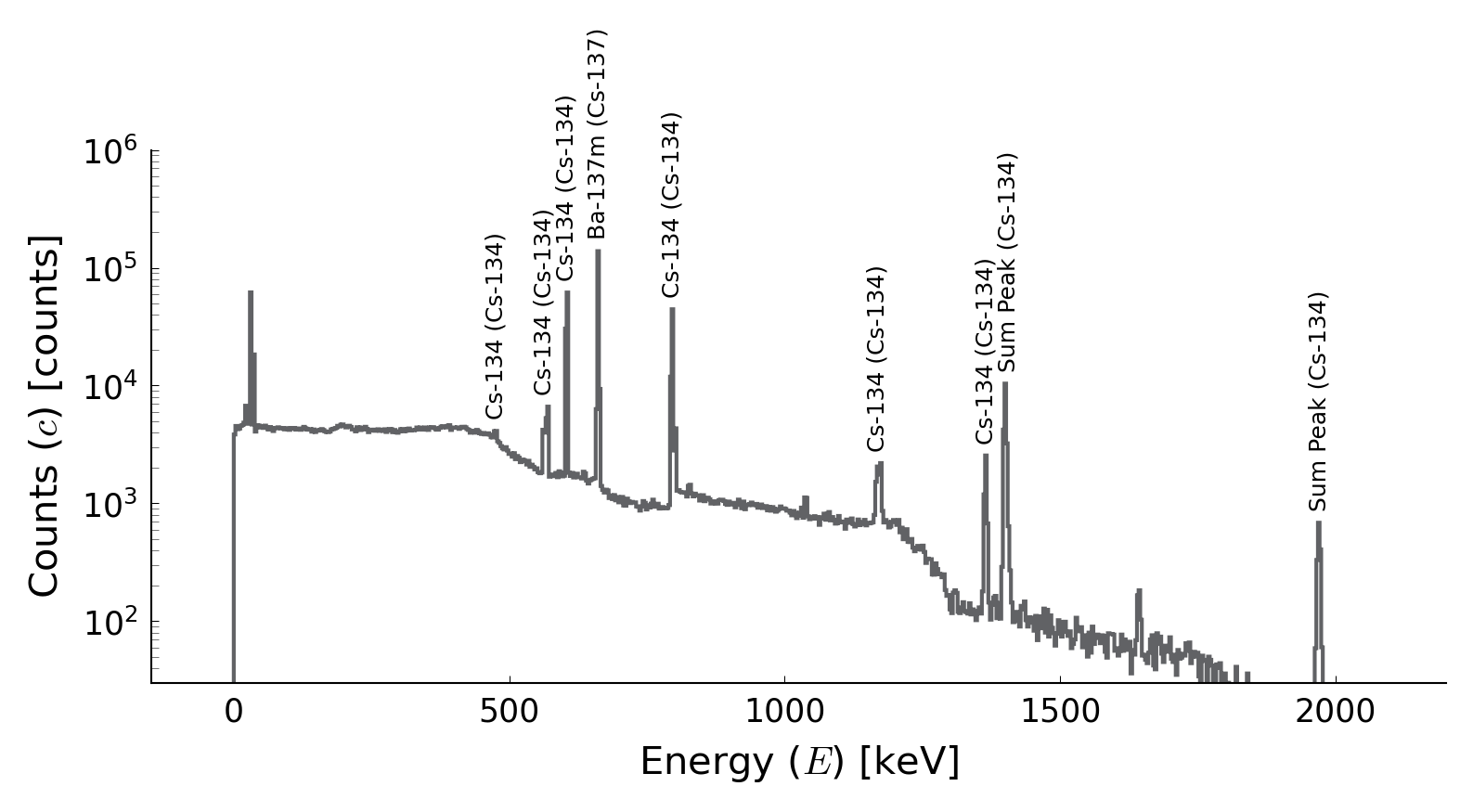}
    \caption{Spectra of detection events in an HPGe detector from \csfour{} and \csseven{} background using our simulation methodology modified from \cite{Keillor2018}, with selected peaks used for quantification of \csfour{} and \csseven{} from \cite{Schoetzig1998}.}
    \label{fig:signal-spectrum-with-peaks}
\end{figure*}

The analytical methodology presented in this work has general application to any information associated with detection events beyond pulse height, such as pulse quality, but we focus on the specific addition of exact time information of each detection event. For application of this analysis technique to continuous monitoring system, similar assumptions would need to be made about filter particulate signal and background, however, a non-well geometry (possibly lower energy resolution) detector and the presence of radioxenon background would need to be assumed. As shown later in our results, high background is the regime on which our method most outperforms the baseline, so the benefit in continuous monitoring could be sizeable. Additionally, radioxenon exhibits coincidence decays ($\beta$-$\gamma$ decays within $\mu s$), so if the detector were given beta detection capabilities (such as a thin window scintillation detector), even greater gains might be possible in a continuous monitoring scenario.

\begin{table}
    \centering
    \caption{The percent of events from background (KUT) and signal (Cs) that generate $N$ detections in a 24 hour counting window.}
    \label{tab:ndets_per_event}
    \begin{tabular}{c|cc}
        $N$ & \textbf{KUT} & \textbf{Cs} \\
        \hline
        1 & 91.6\% & 99.1\% \\
        2 & 8.16\% & 0.87\% \\
        3 & 0.28\% & 0.00\% \\
        4 & 0.00\% & 0.00\%
    \end{tabular}
\end{table}

\subsection{Classifier}
We developed a neural network to classify each detection event into background and signal classes. 
The network took in sequences of length $N_{s}$ detections, each with $N_{f}$ features.
In the simulated data presented, energy and time were the only features provided to the network (i.e. $N_{f} = 2$), and our sequence length was set at 1000 detection events\footnote{Note that sequence length is a trade-off between computational cost and amount of context provided to the model.  One thousand events only covers a small amount of time in our simulation, however the cost of the attention step of our network is $\mathcal{O}\left(N_{s}^{2}\right)$, so this relatively small sequence length was chosen.}.
\subsubsection{Architecture}
 The network takes a sequence input denoted $x$ of size $\mathbb{R}^{N_{s}\times N_{f}}$.
 The energy and time were first pre-processed by normalizing such that their domain is between $-1$ and $1$.
 Then, they were multiplied by a vector of frequencies ranging from $10^{-12}$ to $10^{12}$ of width one quarter of the desired number of ``channels'' ($\frac{N_{c}}{4}$).
 Subsequently, they were used as the argument to  $\sin{()}$ and $\cos{()}$ functions and concatenated, resulting in a frequency domain transform of the input with size $N_{s}\times N_{c}$ which we call the spatio-temporal encoding.
 
 The un-transformed input was used as the input to a five layer multilayer perceptron with $N_{f}$ input nodes and $N_{c}$ nodes per layer thereafter. 
 The output of that neural network was added to the spatio-temporal encoding.

Then, the sum of the transformed input and spatio-temporal encoding was passed through another five layer multilayer perceptron with $N_{c}$ nodes per layer.

\subsubsection{Training}
We train the neural network following standard best practices for supervised training. Our specific settings are:
\begin{description}
    \item[Loss Function] Weighted Cross Entropy loss with accumulating gradients across 4 iterations.
    \item[Iteration size] 1,000 detection events per iteration
    \item[Number of  Iterations] 25,000
    \item[Optimizer] Adam \cite{kingma2014adam} with base learning rate of $1\times 10^{-4}$ and Cosine annealing learning rate scheduler.
\end{description}

The trained model outputs a score in the range $\left[0,1\right]$ indicating how \textit{signal-like} the input event is.
For classifying events, we select the optimal threshold for distinguishing signal vs. background for classifying by choosing the threshold which minimizes the MDA for the assumed background rate in the given context.

\subsection{Baseline Method}

\begin{figure}
    \centering
    \includegraphics{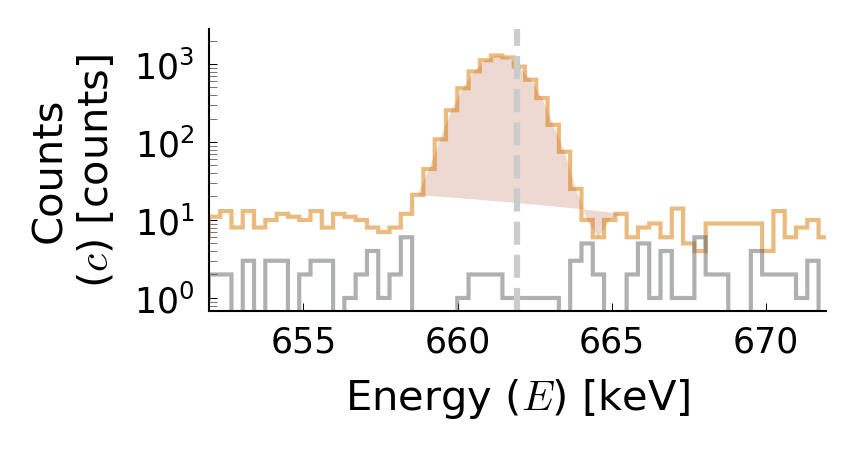}
    \caption{Illustration of the baseline method on the $661\;\mathrm{keV}$ peak of the signal spectrum. The peak is identified from nuclear data sheets, and then two points at several standard deviations (as computed from the detector resolution) on each side of that peak are chosen.  Then, any count in between those two energies in excess of the average number of detections at those two energies is labeled as signal.}
    \label{fig:baseline_method}
\end{figure}

To compare our methods against relevant methods in the gamma spectroscopy field, we selected the direct summation method as our baseline, which is a widely adopted approach for peak detection and quantification, as described in many references including \cite{Knoll2003}.  Several peaks were chosen relevant to the signal isotopes and their major gamma emissions tabulated.  For \csfour{} and \csseven{} these were the $605\;\mathrm{keV}$, $796\;\mathrm{keV}$, and $661.94\;\mathrm{keV}$ peaks.  For each peak energy $E_{pk}$, any detection that was within the signal region, defined as $4$ standard deviations (as computed from the detector resolution function) from $E_{pk}$, was a detection candidate.  Then, the average number of counts in the continuum under that peak was determined by averaging the number of counts at each end point of the signal region.  A detection was classified as signal if it was within the signal region, and was in excess of the average number of counts in the baseline\footnote{Note that the direct summation approach quantifies the peak and does not directly classify each detection directly. Therefore, it is ambiguous as to which detections each ''signal`` label is applied and could have multiple different accuracies for the same minimum detectable activity. The implementation described above is the most generous implementation - it will have the highest classification accuracy for a given peak quantification.}.  A schematic of this method is shown in \cref{fig:baseline_method}.

\section{Results}

Our \texttt{ContextModel} was very successful in performing \csfour{} and \csseven{} detection over the KUT background.  It was able to achieve up to $75\%$ balanced accuracy, which was much higher than that of the baseline method, which achieved approximately $58\%$ balanced accuracy.  The baseline method performed very well against false positives, with $0\%$ false positives reported, but only achieved approximately $10\%$ signal efficiency.  The \texttt{ContextModel} achieved close to $0\%$ false positive rate (though never quite achieving truly zero), while keeping a signal efficiency close to $25\%$.

\begin{figure}
    \centering
    \begin{subfigure}{\columnwidth}
    \includegraphics{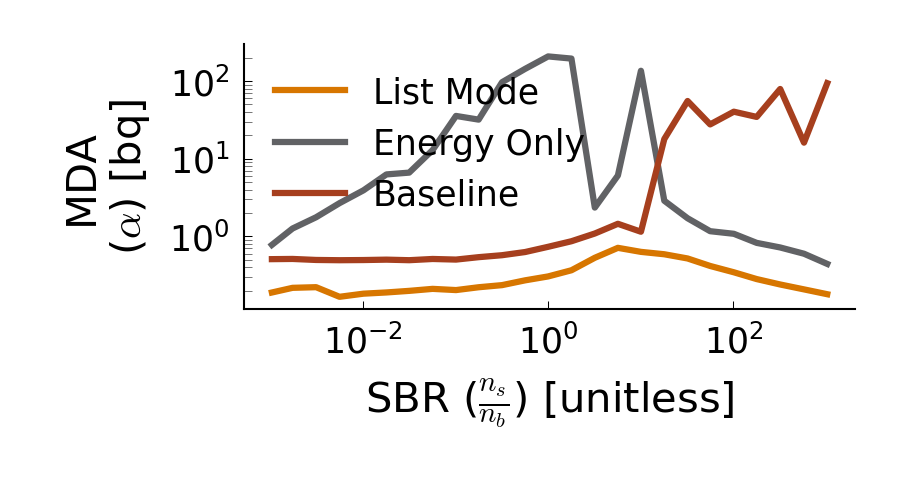}
    \caption{Minimum Detectable Activity (MDA) in bq versus Signal to Background Ratio (SBR) for our List-Mode Method, an Energy Only method, and the baseline method.}
    \end{subfigure}
    \begin{subfigure}{\columnwidth}
    \includegraphics{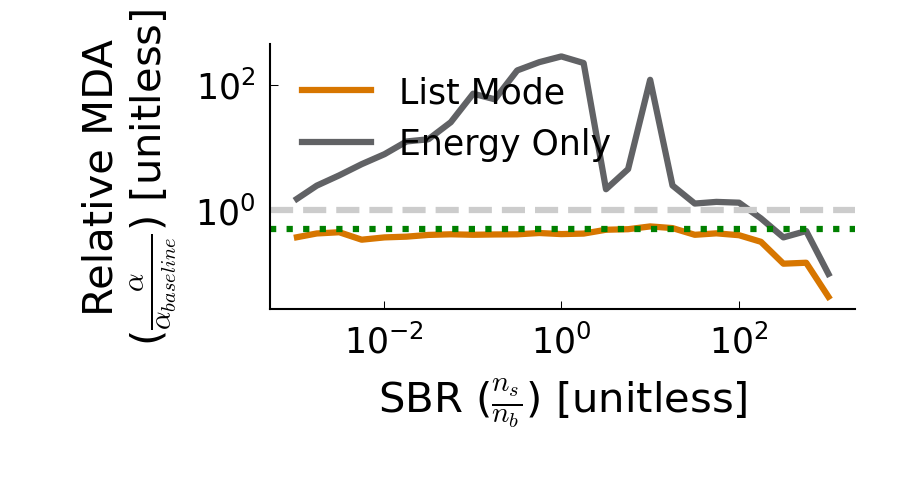}
    \caption{Ratio of the MDA for our List-Mode method and the Energy only Method divided by the MDA of the baseline method versus Signal to Background Ratio (SBR).}
    \end{subfigure}
    \caption{Performance of different classifier based counting experiment methods over a wide range of signal to background ratios.}
    \label{fig:mda_performance}
\end{figure}

The figure of merit for the \texttt{ContextModel} was therefore much better than that for the baseline method.  A signal to background ratio (SBR) must be assumed to calculate a minimum detectable activity (MDA) and that SBR also effects the threshold used to convert \texttt{ContextModel}'s continuous output into a label of ''signal`` or ''background``.  When the SBR is very low, the baseline model performs well due to its low false positive rate, although \texttt{ContextModel}'s higher signal efficiency results in an approximately $2\times$ improvement in MDA for \texttt{ContextModel} compared to the baseline method.  Both the \texttt{ContextModel} and baseline methods performance degrades as SBR increases until an SBR of approximately 10, at which point the baseline model's performance degrades rapidly due to its low signal efficiency, whereas the \texttt{ContextModel}'s performance starts to improve again.  This behavior is shown on \cref{fig:mda_performance}. An energy only model (which does not take into account rate like the baseline method) is also provided for reference, showing worse performance at low SBR, but improving at very high SBR.

\section{Discussion}

\begin{figure*}
    \centering
    \includegraphics{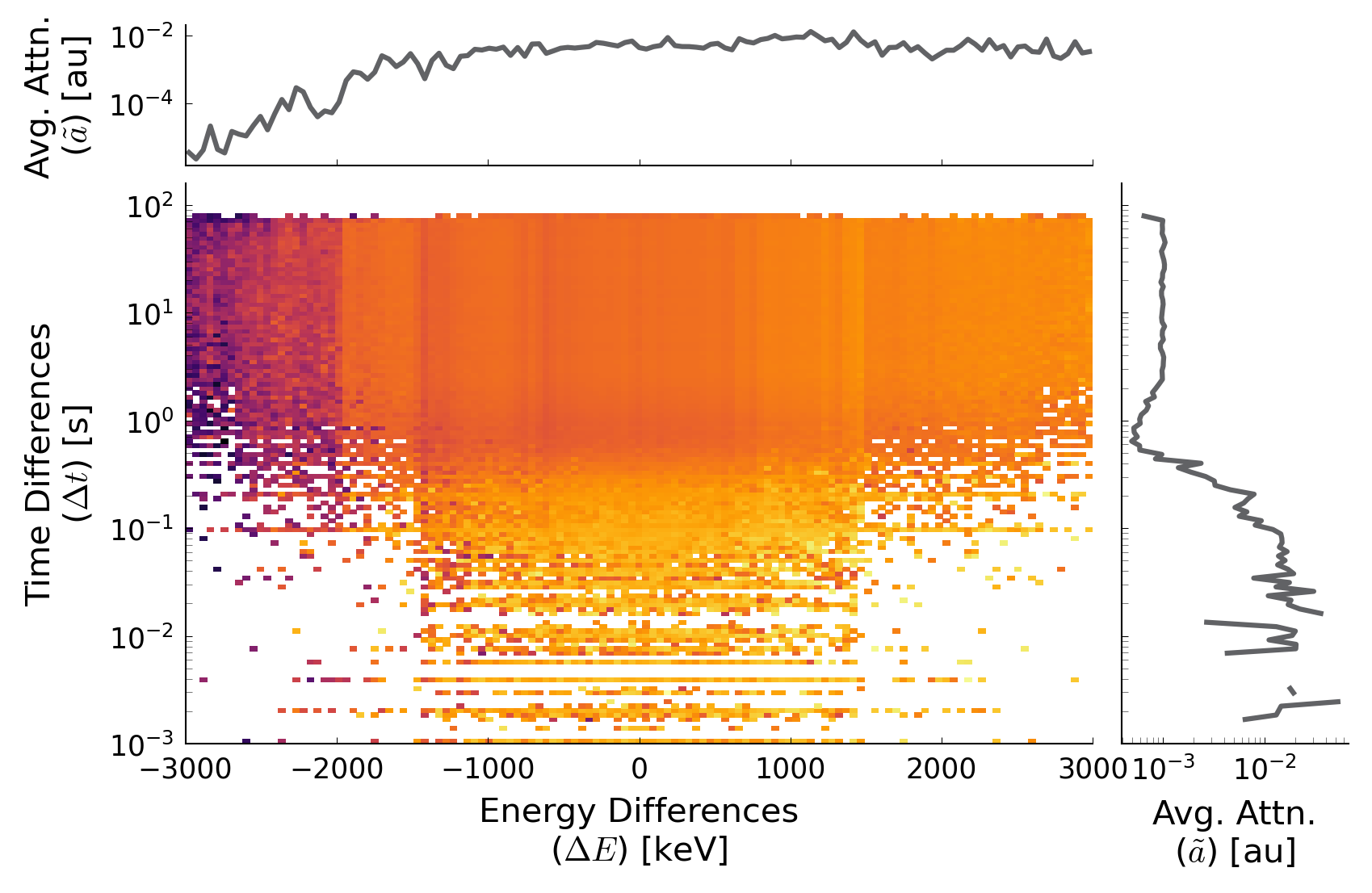}
    \caption{Average attention value for pairs of events versus the difference in energy and logarithm of the absolute value of the time difference between the two detection events. The color indicates the average attention value, with yellow indicating a higher attention value.}
    \label{fig:attention}
\end{figure*}

Thus far, we have presented a model trained to classify gamma spectroscopic detections designed to utilize the energy of detection and timing connection between multiple detection events to generate a detection-by-detection classification of signal or background.  The mechanism we used to allow the connection between multiple detection events is an ``attention mechanism'', which allows for a learned weighted combination of information from all other detection events in a sequence for the classification of a single event. We plot a visualization of the average attention value for pairs of events (indicated by color where yellow indicates higher weighting) compared to the energy difference and the logarithim of the absolute value between the two times. This is shown in \cref{fig:attention}. 

\begin{figure}
    \centering
    \begin{subfigure}{0.5\columnwidth}
    \includegraphics[width=\columnwidth]{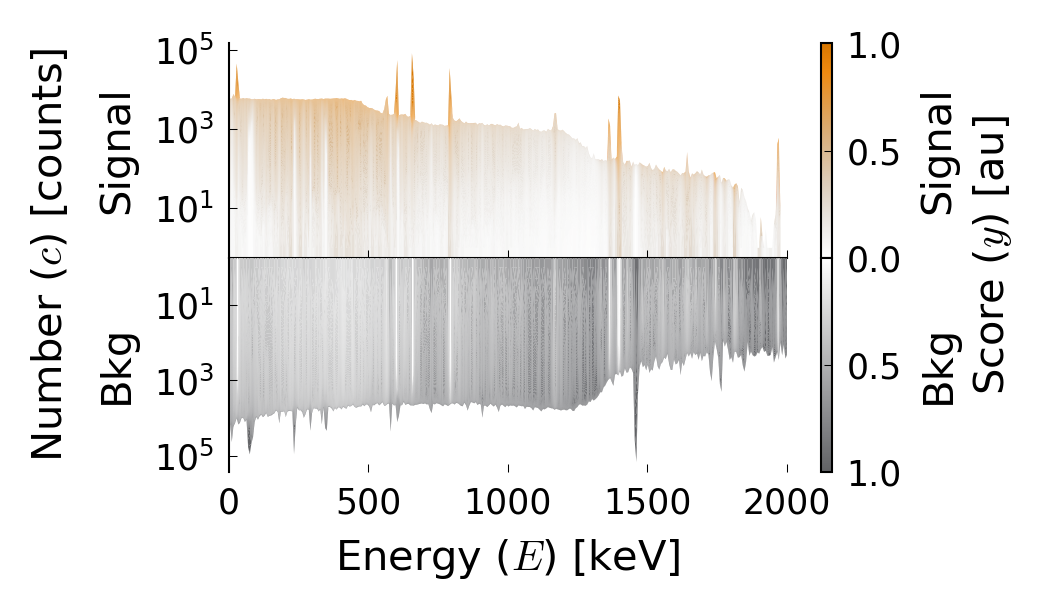}
    \caption{Signal to Background ratio of $\frac{1}{2}$.}
    \end{subfigure}
    \begin{subfigure}{0.5\columnwidth}
    \includegraphics[width=\columnwidth]{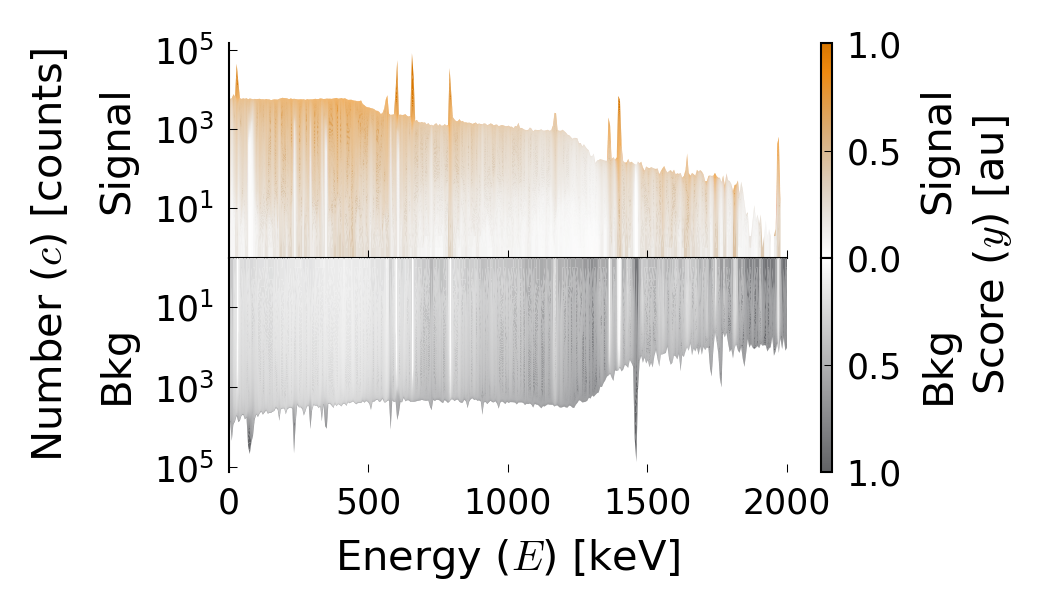}
    \caption{Signal to Background ratio of $1$.}
    \end{subfigure}
    \begin{subfigure}{0.5\columnwidth}
    \includegraphics[width=\columnwidth]{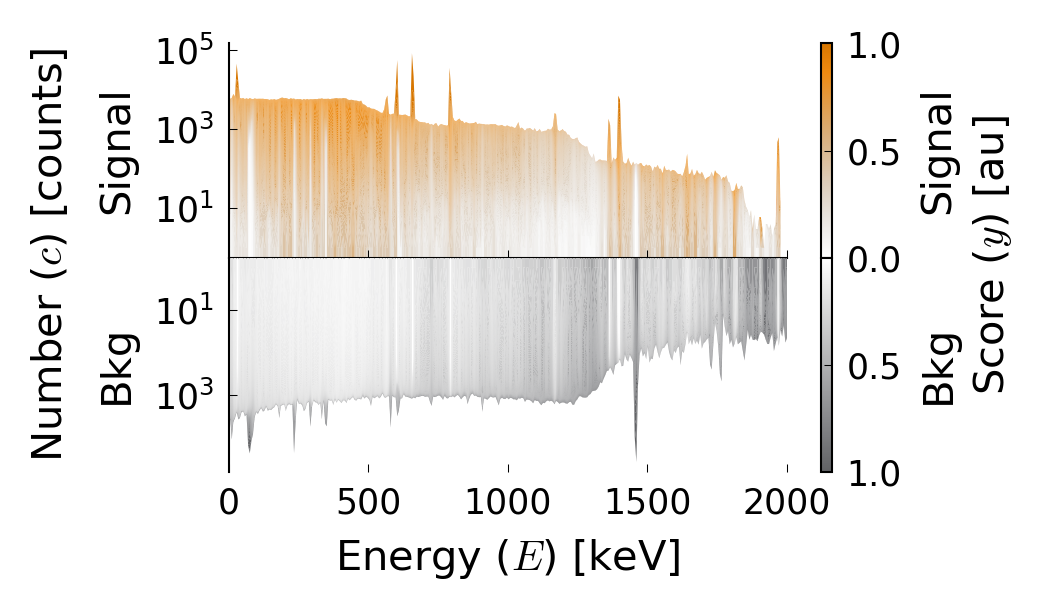}
    \caption{Signal to Background ratio of $2$.}
    \end{subfigure}
    \caption{Signal and Background spectra colored by the signal or background score predicted for each detection event for varying SBRs. The model predicts peaks for each nuclide inventory correctly, and moves on to giving high scores to detections near the compton edge when it is beneficial to the MDA.}
    \label{fig:sbr_vs_spectra}
\end{figure}

In \cref{fig:attention}, the average attention increases with increasing energy difference between the two events, and decreases with increasing time differences between the events.  The energy behavior indicates that learning connections between high energy detections (such as those from Tl-208, which are only present in the background spectrum) and Xrays (which are present in both spectra) are not useful for classification.  The time behavior indicates that long time duration associations between events are less useful than short time associations.  Taken together, \cref{fig:attention} indicates that the \texttt{ContextModel} is likely \emph{not} connecting detection events with past decays of their radioactive predecessors.  Instead, we plot the classification score ($\in \left[0,1\right]$) for signal and background versus differening SBRs in \cref{fig:sbr_vs_spectra}.

In \cref{fig:sbr_vs_spectra}, it can be seen that the most highly scored events for both signal and background are in the peaks that are commonly associated with isotopes known to be in that spectra (e.g. in the 667 keV peak for Cs).  However, as the SBR changes, the \texttt{ContextModel} scores events in the ``compton edge'' more highly, depending if its in a regime where it should weight signal or background more highly.  In the high background regime, when it is beneficial to classify more events as signal, the \texttt{ContextModel} weights heavily those events in the compton edge below the main signal peaks; vice versa for background events in the lower background regime.

Taken together \cref{fig:attention} and \cref{fig:sbr_vs_spectra} show that the model, while not explicitly connecting events with events from their radioactive predecessors or daughters, is using close-time events to determine the overall rate of background events.  Then, it adjusts its predictions based on the most beneficial behavior for the (internally approximated) background rate. The use of the compton edge shows that, while no explicit physics were provided to the model, it was able to learn physically relevant features within the spectra.  While the model did not use our intended mechanism for its predictions, it nonetheless provides performance well beyond the baseline spectroscopy method, enabling $>2\times$ lower minimum detectable activities.

\section{Future Work}

There are many opportunities for extension of this framework and its application to other detection problems.  In particular, the theoretical backing behind the framework would be strengthened by investigation into those locations where an improvement in mutual information does not lead to an improvement in the figure of merit, by exploration of those physical systems which can or cannot take advantage of time resolved spectroscopy (such as all singles detection systems), and by the use of weakly supervised frameworks for training without simulated data.  Many applications also exist, such as for the detection of transient isotopic content in gasses in radioxenon detection, for the detection of activation products or photofissions in active interrogation systems, or for the blanking of high intensity photons in active interrogation systems. Ultra-low background systems are also ripe for exploitation of this framework, especially those with instrumentation for extra features about each detection event, such as electrical pulse quality. Finally, investigations using longer event sequence lengths would be beneficial, and could be achieved by utilizing more memory-efficient approximate attention implementations to get around the computational barrier posed by the standard $\mathcal{O}\left(N_{s}^{2}\right)$ implementation of the attention mechanism.

\section{Acknowledgements}

This work was supported by the Office of Defense Nuclear Nonproliferation Research and Development within the U.S. Department of Energy’s National Nuclear Security Administration and Pacific Northwest National Laboratory, which is operated by Battelle Memorial Institute for the U.S. Department of Energy under contract DE-AC05-76RL01830.

\bibliographystyle{elsarticle-num}
\bibliography{references.bib}

\begin{appendices}

\section{Proof: Additional input feature cannot decrease the mutual information}\label{sec:proof-feature-mi-increase}
The mutual information between features $X$ and labels $Y$ is, in the continuous case

\begin{equation}\label{eq:def-mi}
\begin{aligned}
I\left(X;Y\right) =&
\int_{\mathcal{Y}}\int_{\mathcal{X}}P_{\left(X,Y\right)}\left(x,y\right)\cdot 
\\
&\quad \quad \log\left(\frac{P_{\left(X,Y\right)}\left(x,y\right)}{P_{X}\left(x\right)P_{Y}\left(y\right)}\right)dxdy.
\end{aligned}
\end{equation}

We explore the case for traditional spectroscopy, where the input features are solely the energy $X \equiv E$, versus generalized graph based spectroscopy, where $X \equiv \left[E, t, \dots\right]$. Therefore, the two integrals become

\begin{equation}\label{eq:mi-spectroscopy}
\begin{aligned}
I\left(E;Y\right) =& \int_{\mathcal{Y}}\int_{D\left(E\right)}P_{\left(E,Y\right)}\left(\varepsilon,y\right)\cdot 
\\
& \quad \quad \log\left(\frac{P_{\left(E,Y\right)}\left(\varepsilon,y\right)}{P_{E}\left(\varepsilon\right)P_{Y}\left(y\right)}\right)d\varepsilon dy,
\end{aligned}
\end{equation}
where $D\left(E\right)$ is the domain of all possible energies, and

\begin{equation}\label{eq:mi-graph-spectroscopy}
\begin{aligned}
I\left(\left[E, t\right];Y\right) =&  \int_{\mathcal{Y}}\int_{D\left(E\right)}\int_{D\left(t\right)}P_{\left(\left[E, t\right],Y\right)}\left(\left[\varepsilon, \tau\right],y\right)\cdot 
\\
& \log\left(\frac{P_{\left(\left[E, t\right],Y\right)}\left(\left[\varepsilon, \tau\right],y\right)}{P_{\left[E, t\right]}\left(\left[\varepsilon, \tau\right]\right)P_{Y}\left(y\right)}\right)d\tau d\varepsilon dy,
\end{aligned}
\end{equation}
where $D\left(t\right)$ is the domain of all possible times.

From the definition of marginal probability,
\begin{equation}\label{eq:pey-marginal}
P\left(E,Y\right) \equiv \int_{D\left(t\right)} P_{\left(\left[E, t\right], Y\right)}\left(\left[\varepsilon, \tau\right], y\right) d\tau,
\end{equation}
and
\begin{equation}\label{eq:pe-marginal}
P_{E}\left(\varepsilon\right) \equiv \int_{D\left(t\right)} P_{\left[E, t\right]} \left(\left[\varepsilon, \tau \right]\right)d\tau.
\end{equation}
These values are by definition independent of $\tau$, which leads to the fact that
\begin{equation}\label{eq:dpey}
    \frac{\partial P_{\left(E, Y\right)}}{\partial \tau} = 0
\end{equation} and 
\begin{equation}\label{eq:dpe}
    \frac{\partial P_{E}}{\partial \tau} = 0.
\end{equation}
We can move the term to the left of the minus sign in \cref{eq:mi-spectroscopy} inside an integral with respect to $d\tau$ by defining quantities $\tilde{P}_{\left(\left[E,t\right],Y\right)}$ and $\tilde{P}_{\left[E,t\right]}$ which satisfy \cref{eq:pey-marginal,,eq:pe-marginal,,eq:dpey,,eq:dpe}, giving
\begin{equation}\label{eq:mi-spectroscopy-expanded}
\begin{aligned}
I\left(E;Y\right) &=
\int_{\mathcal{Y}}\int_{D\left(E\right)}\int_{D\left(t\right)}\tildepety 
\\
& \log \left(\frac{\tildepety}{\tildepet \pey}\right) d\tau d\varepsilon dy,
\end{aligned}
\end{equation}

We note that both \cref{eq:mi-spectroscopy,,eq:mi-graph-spectroscopy} are of the form
\begin{equation}
    h\left(a, b, c\right) = a\log\left(\frac{a}{bc}\right),
\end{equation}
where $a$ and $b$ may be functions of $\tau$. The Hessian of $h$ with respect to $a$ and $b$ is then
\begin{equation}
    \mathbb{H}\left(h\left(a, b\right)\right) = \left[
    \begin{array}{cc}
        \frac{1}{a} & -\frac{1}{b} \\
        -\frac{1}{b} & \frac{a}{b^{2}}
    \end{array}
    \right]
\end{equation}
which is symmetric and has eigenvalues
\begin{equation}
    \lambda_{1} = 0,
\end{equation}
and
\begin{equation}
    \lambda_{2} = \frac{a^{2} + b^{2}}{ab^{2}},
\end{equation}
which are both non-negative. Therefore $f\left(a, b, c\right)$ is convex with respect to $a$ and $b$, therefore
\begin{equation}\label{eq:convexity}
    f\left(\mathbb{E}\left(a\right), \mathbb{E}\left(b\right), c\right) \geq \mathbb{E}_{a,b}\left(f\left(a, b, c\right)\right).
\end{equation}
The expectation value of $\pety$ is
\begin{equation}
\begin{aligned}
    \mathbb{E}\left(\pety\right) = \\ 
    \frac{1}{V\left(t\right)}\int_{D\left(t\right)} \pety d\tau,
\end{aligned}
\end{equation}
and the expectation value of $\pet$ is 
\begin{equation}
    \mathbb{E}\left(\pet\right) = \frac{1}{V\left(t\right)}\int_{D\left(t\right)} \pet d\tau,
\end{equation}
where $V\left(t\right)$ is the volume of the domain of $t$. Then, from \cref{eq:mi-spectroscopy-expanded},
\begin{equation}\label{eq:i-ey-expect}
\begin{aligned}
    I\left(E;Y\right) = \int_{\mathcal{Y}}\int_{D\left(E\right)} V\left(t\right)\\
    \mathbb{E}_{a,b}\left(f\left(\pety, \pet,\right.\right.\\
    \left.\left.\pey\right)\right)d\varepsilon dy,
\end{aligned}
\end{equation}
and from \cref{eq:mi-graph-spectroscopy},
\begin{equation}\label{eq:i-ety-expect}
\begin{aligned}
    I\left(\left[E, t\right];Y\right) = \int_{\mathcal{Y}}\int_{D\left(E\right)}V\left(t\right)\\
    f\left(\mathbb{E}\left(\pety\right),\mathbb{E}\left(\pet\right),\right. \\\left.\pey\right) d\varepsilon dy.
\end{aligned}
\end{equation}

The integrand in \cref{eq:i-ey-expect} is clearly bounded above by that in \cref{eq:i-ety-expect} from the property of convexity in \cref{eq:convexity}. Therefore, the overall integral in \cref{eq:i-ey-expect} is bounded above by the overall integral in \cref{eq:i-ey-expect}. This proves that the mutual information between simple spectroscopic data must always be equal or less than that in graph spectroscopic data. By defining $t$ and its domain $D\left(t\right)$ generally, this proof can be extended to additional information of any type or dimension.

\section{Proof: TJPs Must Vary}\label{sec:tjp-must-vary}
The theoretical result in the previous section gives us strong motivation for usage of our graph formulation, or indeed to add as much information as possible to any detection experiment analysis; but that neglects the physical realities and equipment limitations in actual spectroscopy.  However, \cref{eq:mi-graph-spectroscopy,,eq:mi-spectroscopy-expanded} give us the ability to determine actual limits for what isotopes and situations might actually give rise to improved performance. Improved performance requires that $I\left(E, Y\right) < I\left(\left[E, t\right], Y\right)$, or equivalently that $I\left(E, Y\right) - I\left(\left[E, t\right], Y\right) < 0$.  This difference, after rearrangement with some properties of the logarithm, can be written
\begin{multline}
    \Delta = \int_{\mathcal{Y}}\int_{D\left(E\right)}\int_{D\left(t\right)} \\
    \tildepety \log \left(\tildepety\right)\\
    - \tildepety \log \left(\tildepet \right) \\
    - \pety \log \left( \pety \right)\\
    + \pety \log \left( \pet \right) \\
    - \tildepety \log \left( \pey \right) \\
    + \pety \log \left( \pey \right) \\d\tau d\varepsilon dy.
\end{multline}

Separating the last two terms of the integrand, and noting that they cancel out over the integration with respect to $\tau$ due to the definition of $\tildepety$, we are interested in the physically realistic situations where
\begin{multline}\label{eq:mi-diff}
    \int_{\mathcal{Y}}\int_{D\left(E\right)}\int_{D\left(t\right)} \\
    \tildepety \log \left(\tildepety\right) \\
    - \tildepety \log \left(\tildepet \right) d\tau d\varepsilon dy \\
    < \int_{\mathcal{Y}}\int_{D\left(E\right)}\int_{D\left(t\right)} \\
    \pety \log \left( \pety \right) \\
    - \pety \log \left( \pet \right) d\tau d\varepsilon dy.
\end{multline}

For \cref{eq:mi-diff} to be true, $\pety$ and $\pet$ (hereafter temporal joint probabilities (TJPs)) must differ from $\tildepety$ and $\tildepet$. Because of the definitions of $\tildepety$ and $\tildepet$, that means that t
\end{appendices}
\end{document}